# Period doubling eigenstates in a fiber laser mode-locked by nonlinear polarization rotation

XIONGQUAN YAO,[1] LEI LI,[1,*] ANDREY KOMAROV,[2] MARIUSZ KLIMCZAK,[3,4] DINGYUAN TANG,[5] DEYUAN SHEN,[1] LEI SU,[6] AND LUMING ZHAO,[1,6,*]

[1]*Jiangsu Key Laboratory of Advanced Laser Materials and Devices, Jiangsu Collaborative Innovation Center of Advanced Laser Technology and Emerging Industry, School of Physics and Electronic Engineering, Jiangsu Normal University, Xuzhou, 221116 Jiangsu, P. R. China*
[2]*Institute of Automation and Electrometry, Russian Academy of Sciences, Academician Koptyug Prospekt 1, 630090 Novosibirsk, Russia*
[3]*Faculty of Physics, University of Warsaw, Warsaw, Poland*
[4]*Institute of Electronic Materials Technology, Wólczynska 133, 01-919 Warsaw, Poland*
[5]*School of Electrical and Electronic Engineering, Nanyang Technological University, Singapore 639798*
[6]*School of Engineering and Materials Science, Queen Mary University of London, London E1 4NS, UK*
*\* sdulilei@gmail.com, lmzhao@ieee.org*

**Abstract:** Due to the weak birefringence of single mode fibers, solitons generated in fiber lasers are indeed vector pulses and exhibit periodic parameter change including polarization evolution even when there is a polarizer inside the cavity. Period doubling eigenstates of solitons generated in a fiber laser mode-locked by the nonlinear polarization rotation, i.e., period doubling of polarization components of the soliton, are numerically explored in detail. We found that, apart from the synchronous evolution between the two polarization components, there exists asynchronous development depending on the detailed operation conditions. In addition, period doubling of one polarization component together with period-one of another polarization component can be achieved. When the period tripling window is obtained, much complexed dynamics on the two polarization components could be observed.

## 1. Introduction

Period doubling bifurcation in mathematics refers to a bifurcation in which a slight change of a parameter value in a dynamic system's equations results in the system switching to a new behavior with twice the period of the original system. With the doubled period, the system repeats themselves taking twice as many iterations as before. Bifurcation phenomena include period doubling, period tripling, period n (n>3), and period doubling route to chaos etc. Period doubling bifurcation is one of the universal properties of nonlinear systems. Ultrafast Fiber lasers are a paradigm of nonlinear systems. Due to the feasibility of easy controlling of laser parameters, small cross-section of light propagation, and long interaction distance between light and fibers forming the laser, fiber lasers are a perfect platform for studying dynamics in nonlinear systems.

When the dispersion effect balances the nonlinear Kerr effect during pulse propagation, a soliton can be generated in a fiber laser. Period doubling of solitons in fiber lasers was predicted two decades ago [1] and soon be demonstrated [2]. Period doubling, period tripling, and the corresponding bifurcations of solitons have been reported [3-11]. Based on the classical one-dimension quantic complex Ginzburg-Landau equation (GLE) and the varying of the nonlinear gain, period doubling and period tripling were discovered [3]. By varying spectral filtering or altering group velocity dispersion coefficient of the gain fiber, period tripling bifurcation was numerically observed [4]. The changing of nonlinear gain alone, the varying of spectral filtering alone, or the altering of group velocity dispersion coefficient of the gain fiber alone is difficult to realize in practice. Zhang et al numerically achieved period-tripling bifurcation route to chaos as well as the classical period-doubling bifurcation route to chaos by changing the small signal gain only [5]. It simplifies the theme that is capable to experimentally demonstrate period-tripling bifurcation by changing pump level only. Period doubling and its variants are intensity threshold dependent. The appearance of period doubling or period-doubling bifurcations is subject to that the accumulated nonlinearity during pulse propagation is strong enough. Therefore, period doubling is independent of the cavity dispersion of fiber lasers. So far, period doubling has been demonstrated in fiber lasers of either anomalous dispersion [6], normal dispersion [7], or around zero dispersion [8]. Research progress on period doubling bifurcation in fiber lasers is well reviewed recently [9].

Due to the bias from the theoretically symmetrical cross-section of fibers and the random operation conditions, fibers have at least weak birefringence. Consequently, solitons generated in a fiber laser are vector solitons even when the fiber laser includes polarization sensitive components, for example, a polarizer. Previously, solitons, generated in fiber lasers mode-locked by using the nonlinear polarization rotation (NPR) technique [6-8], were considered as scalar solitons. Consequently, the period doubling phenomena were analyzed based on single polarization even the fiber laser was described by coupled GLEs. In another word, the period doubling phenomena reported are actually based on the solitons composed of two polarization components. It is a composite state instead of the eigenstates. Wu et al have demonstrated both numerically and experimentally that vector and scalar solitons can co-exist within the laser cavity in a passively mode-locked fiber laser by using the NPR technique [10]. The mode-locked pulse evolves as a vector soliton in the strong birefringent segment and is transformed into a scalar soliton after the polarizer in the laser cavity. Period doubling of vector solitons has been reported a decade ago, where there is no polarization dependent component in the fiber laser and the mode locking mechanism is also polarization-independent. The two components of the period-doubled vector soliton each can exhibit period doubling [11]. The pulse intensity variation between the two orthogonal polarization components could be anti-phase, which results in weak period doubling (small pulse intensity shuttle) of the vector soliton. Hence, it is interesting and necessary to comprehend the detailed period doubling performance of polarization components of a soliton generated from a fiber laser mode-locked by using the NPR technique. In this paper, we numerically studied period doubling eigenstates in a fiber laser mode-locked by using the NPR technique. When period doubling of a soliton is achieved, apart from the synchronous evolution between the two polarization components of the soliton, we observed asynchronous development depending on the detailed operation conditions. In addition, period doubling of one polarization component together with period-one of another polarization component could be obtained. Much complexed dynamics on the two polarization components could be observed when the period tripling window is achieved.

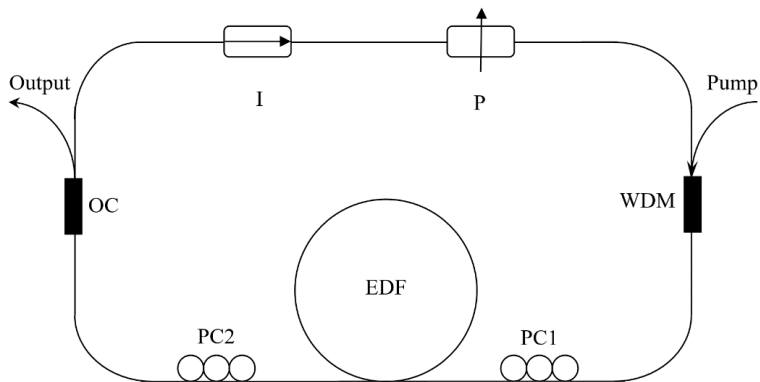

Fig. 1 Schematic of the fiber laser. WDM: wavelength division multiplexer; PC1/PC2: polarization controller; EDF: Erbium-doped fiber; OC: output coupler; I: isolator; P: polarizer.

## 2. Fiber laser setup and theoretical model

Numerical simulations were carried out based on the fiber laser shown in Fig. 1. A polarizer and two polarization controllers (PCs) were used to employ the NPR technique for achieving mode locking. An isolator was used to guarantee the unidirectional operation. Roundtrip model was taken to describe pulse evolution in the laser [12]. In general, an arbitrary small pulse is used as the initial input and it circulates in the laser cavity until a steady pulse evolution is established under specific operation conditions. When the pulse goes through a component, the Jones matrix of the component is applied to the pulse. The pulse propagation in fibers is described by the coupled GLEs as shown in equation (1), where the dispersion effect, nonlinear effect, gain and gain dispersion effect are included:

$$\begin{cases} \dfrac{\partial u}{\partial z} = i\beta u - \delta \dfrac{\partial u}{\partial t} - \dfrac{ik''}{2}\dfrac{\partial^2 u}{\partial t^2} + \dfrac{ik'''}{6}\dfrac{\partial^3 u}{\partial t^3} + i\gamma(|u|^2 + \dfrac{2}{3}|v|^2)u + \dfrac{i\gamma}{3}v^2 u^* + \dfrac{g}{2}u + \dfrac{g}{2\Omega_g^2}\dfrac{\partial^2 u}{\partial t^2} \\ \dfrac{\partial v}{\partial z} = -i\beta v + \delta \dfrac{\partial v}{\partial t} - \dfrac{ik''}{2}\dfrac{\partial^2 v}{\partial t^2} + \dfrac{ik'''}{6}\dfrac{\partial^3 v}{\partial t^3} + i\gamma(|v|^2 + \dfrac{2}{3}|u|^2)v + \dfrac{i\gamma}{3}u^2 v^* + \dfrac{g}{2}v + \dfrac{g}{2\Omega_g^2}\dfrac{\partial^2 v}{\partial t^2} \end{cases} \quad (1)$$

where u and v are the complex optical envelopes in orthogonal polarization mode along fibers; $u^*$ and $v^*$ are the conjugates of u and v; $\beta = \pi\Delta n/\lambda$ is the wave-number difference between the two polarization modes of the fiber, where $\Delta n$ is the difference between the effective indices of the two modes, and $\lambda$ is the wavelength; $\delta = \beta\lambda/2\pi c$ is the inverse of group velocity dispersion, where c is the light speed; $k''$ is the second order dispersion coefficient, $k'''$ is the third order dispersion coefficient, and $\gamma$ represents the nonlinearity of the fiber. g is the saturable gain coefficient of the gain fiber and $\Omega_g$ is the gain bandwidth. For undoped fibers g = 0; for EDF, the gain saturation is:

$$g = G\exp[-\dfrac{\int(|u|^2+|v|^2)dt}{P_{sat}}] \quad (2)$$

To study period doubling eigenstates, the following parameter set is selected as shown in Table 1.

Table 1. Parameters used in the simulations [12]

| $\gamma$ | 3 W$^{-1}$km$^{-1}$ | $k''_{EDF}$ | -23 ps$^2$/km |
|---|---|---|---|
| $k''_{SMF}$ | -23 ps$^2$/km | $k'''$ | -0.13 ps$^3$/km |
| $L_b$ | 6 m | $\Omega_g$ | 50 nm |
| $\theta$ | $0.152\pi$ | $\varphi$ | $\theta+\pi/2$ |
| $P_{sat}$ | 250 | G | varying |
| $\Phi_{PC}$ | varying between $\pi$ to $2\pi$ (mode locking regime) | | |

where $L_b$ is the birefringence beat length, and the laser cavity length L = 4 m, which is composed of 1 m SMF, 1 m EDF, and 2 m SMF in sequence, and the output position is at the end of the first segment of SMF, and 10% output is applied. The definition of other parameters can be retrieved from Ref. [12]

## 3. Simulation results

By properly choosing the linear cavity phase delay bias, which corresponds in the experiment to appropriately selecting the orientations of the polarization controllers, soliton operation can be always obtained in our simulations. As the value of gain increases, the period-doubling state can be achieved. But with different linear cavity phase delay bias, each polarization component of the soliton can show individual behavior of period doubling.

When the linear cavity phase delay bias is set at $1.35\pi$, a stable soliton sequence can be obtained when the gain coefficient is set at $G = 700$, as shown in Fig. 2(a). Increasing the value of $G$ and keeping all the other parameters fixed, the soliton repetition period in the cavity is doubled at $G = 800$ (Fig. 2(c)). Figure 2(b) and Fig. 2(d) shows the corresponding optical spectrum of the solitons to Fig. 2(a) and Fig. 2(c), where the optical spectrum of 10 roundtrips are overlapped. Therefore, the spectra shown in Fig. 2(d) have two different curves corresponding to period doubling while the spectra shown in Fig. 2(b) have only one curve.

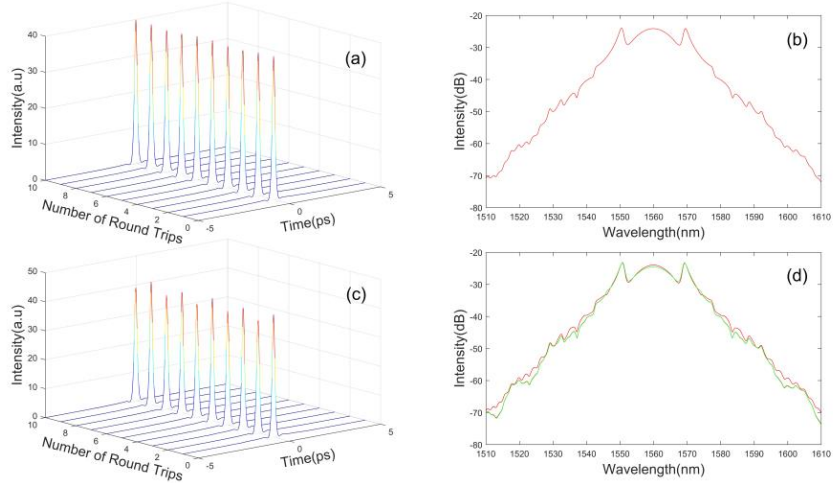

Fig. 2 Soliton pulse evolution and the corresponding optical spectra numerically calculated under different pump strength. The linear cavity phase delay bias is set as $\Phi_{PC} = 1.35\pi$. (a)/(b) Period-one soliton state, $G = 700$; (c)/(d) Period-two soliton state, $G = 800$

To study the detailed period doubling performance of the two orthogonal polarization components, the gain coefficient is set at $G = 820$ to make the cavity operated in a period-doubling state and change the linear cavity phase delay bias to understand the pulse intensity variation between horizontal component and vertical component, as shown in Fig. 3. When $\Phi_{PC} = 1.35\pi$, the pulse train of horizontal component and vertical component is shown in Fig. 3(a) and Fig. 3(b). The evolution of horizontal component and vertical component are synchronous. When the linear cavity phase delay bias changes to $\Phi_{PC} = 1.42\pi$, the horizontal component becomes period-one state while the vertical component is period-doubling state, as shown in Fig. 3(c)/(d). Changing $\Phi_{PC}$ to $1.46\pi$, the horizontal component and vertical component are all in period-two states but they are asynchronous, as shown in Fig. 3(e)/(f).

When the NPR technique is used for achieving mode locking in fiber laser, the generated ultrashort pulse is generally considered as a scalar soliton as the pulse polarization is periodically normalized to the polarization direction of the polarizer. However, due to the existence of fiber weak birefringence, the soliton obtained in the fiber laser indeed consist of two orthogonal polarization components, although the soliton is polarization-normalized every time when it passes through the polarizer. Consequently, the period doubling eigenstates are not based on the soliton anymore but should be based on the polarization components of the soliton. Simulation results shown in Fig. 3 clearly suggests that the period doubling of the polarization component could be different from that of the soliton. When the soliton exhibits period doubling, depending on the detailed operation conditions the polarization component could show synchronous period doubling, asynchronous period doubling, or even period one, respectively.

Numerically we also observed period tripling of soliton and the corresponding period tripling of individual polarization component. So far only the synchronous behavior was observed as shown in Fig. 4. We believe that more complexed behaviors apart from the synchronous one exist depending on the detailed operation conditions.

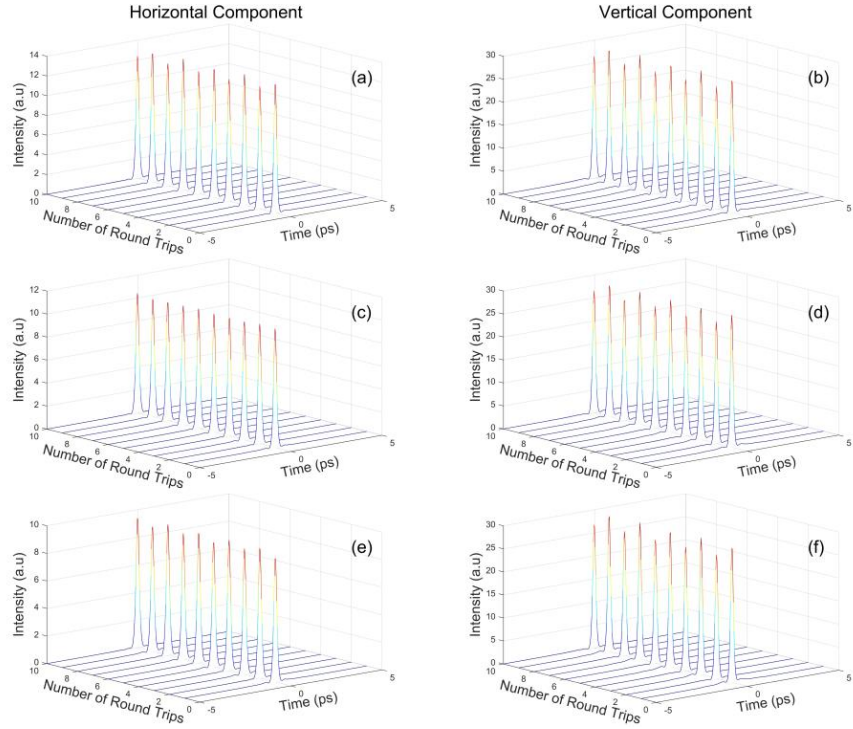

Fig. 3 The detailed period doubling eigenstates of period doubling solitons under different linear cavity phase delay bias. The gain coefficient is fixed at $G = 820$. (a)/(b) synchronous when $\Phi_{PC} = 1.35\pi$; (c)/(d) the horizontal component is period one while the vertical component is period doubling when $\Phi_{PC} = 1.42\pi$; (e)/(f) asynchronous when $\Phi_{PC} = 1.46\pi$

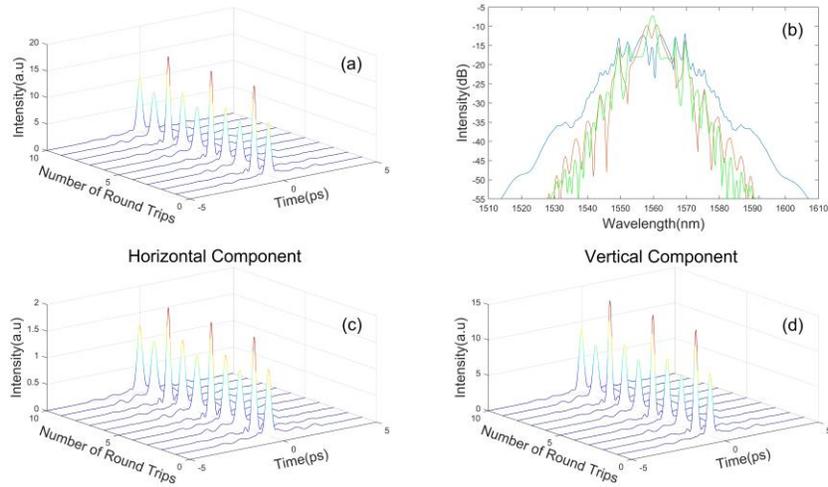

Fig. 4 Period tripling of soliton and the corresponding period tripling of individual polarization component. (a)/(b) Period tripling of soliton and the corresponding optical spectra; (c)/(d) Period tripling of soliton of individual polarization component

## 4. Conclusions

Appearance of period doubling and its variants is one of intrinsic properties of nonlinear systems. Period doubling eigenstates are discussed in a fiber laser mode-locked by using the NPR technique. Numerical simulations suggest that each polarization component of the soliton could show independent behavior of period doubling while there exists interaction between the two orthogonal polarization components and the polarization was normalized every time when the soliton periodically passes the polarizer. Depending on the detailed parameters, either synchronous evolution or asynchronous development between the two polarization components of the generated soliton under period doubling appearance can be achieved. Specifically, period doubling of one polarization component while the other polarization component maintaining period-one can as-well be achieved. The discovery above great enriches our understanding on soliton dynamics in fiber lasers and pave the way to be closer to the real dynamics of nonlinear systems.

## Funding

National Natural Science Foundation of China (NSFC) (11674133, 11911530083, 61575089); Russian Foundation for Basic Research (RFBR) (19-52-53002); Royal Society (IE161214); Protocol of the 37th Session of China-Poland Scientific and Technological Cooperation Committee (37-17); European Union's Horizon 2020 research and innovation programme under the Marie Skłodowska-Curie grant agreement No. 790666. We acknowledge support from Jiangsu Overseas Visiting Scholar Program for University Prominent Young & Middle-aged Teachers and Presidents and Priority Academic Program Development of Jiangsu Higher Education Institutions (PAPD). Mariusz Klimczak acknowledges support from Fundacja na rzecz Nauki Polskiej (FNP) in scope of First TEAM/2016-1/1 (POIR.04.04.00-00-1D64/16). Postgraduate research innovation program of Jiangsu Normal University (2018YXJ596).

## Disclosures

The authors declare that there are no conflicts of interest related to this article.